\documentclass[12pt,draft]{IEEEtran} %!PN
\input{epsf.sty}

\def\BibTeX{{\rm B\kern-.05em{\sc i\kern-.025em b}\kern-.08em
    T\kern-.1667em\lower.7ex\hbox{E}\kern-.125emX}}

\usepackage{ieeefig}

\begin{document}
\title{Role of scattering in nanotransistors}
\author{Alexei Svizhenko and M. P. Anantram\thanks{NASA Ames Research
Center, Mail Stop: 229-1, Moffett Field, CA 94035-1000.}}

\markboth{IEEE Transactions in Electron Devices}
%{Murray and Balemi: Using the style file IEEEtran.sty} %!PN^M
{Svizhenko and Anantram: Using the Document Class IEEEtran.cls} %!PN^M

\maketitle

\begin{abstract}
We model the influence of scattering along the channel and extension 
regions of dual gate nanotransistor. It is found that the reduction in 
drain current due to scattering in the right half of the channel is 
comparable to the reduction in drain current due to scattering in the
left half of the channel, when the channel length is comparable to the
scattering length. This is in contrast to a popular belief that 
scattering in the source end of a nanotransistor is significantly more
detrimental to the drive current than scattering elsewhere. As the 
channel length becomes much larger than the scattering length, 
scattering in the drain-end is less detrimental to the drive current 
than scattering near the source-end of the channel. 
Finally, we show that for nanotransistors, the classical picture of 
modeling the extension regions as simple series resistances is not
valid.

\vspace{2in}

Accepted for publication in "IEEE Transaction in Electron Devices on
Electron Devices"
\end{abstract}

%\narrowtext
\pagebreak

\section{Introduction}
\label{sect:introduction}

Experimental and theoretical work on nanotransistors has been a hot 
area of research because of significant advance in lithography. The 
significant advances in lithography have led to the construction of 
nanotransistors with channel lengths smaller than 25 nanometers 
(nm)~\cite{kedzierski-slm-00,hokazono-iedm-02,boeuf-iedm-01}. It is 
believed that devices with channel lengths equal to 10 nm may become 
possible in research laboratories~\cite{yu-iedm-02}. In these
nanotransistors, the length scales of the channel, gate, screening
and scattering lengths, begin to become comparable to one another. This
is not the case for long channel MOSFETs, where the channel and
gate lengths are much larger than the scattering lengths. As a result
of the comparable length scales, it is expected that the physics of
nanotransistors will begin deviating from that of long channel 
transistors.

The resistance of a MOSFET (Fig. \ref{fig:dg}) with a long channel
length can be qualitatively thought of as arising in four regions, 
Extension regions near the source (Ex-s) and drain (Ex-d), Channel
(Ch), and Contacts. It is believed that the resistance of the contacts
and extension regions are extrinsic series resistances~\cite{taur-book}, 
while the channel resistance is intrinsic to the MOSFET. For a given 
doping distribution, both electrostatics and {\it scattering} of the
current carriers play an important role in determining the drive current.
Electrostatics dictates that the total carrier density in a long 
channel MOSFET is approximately $C_{ox} (V_G-V_S)$, where $V_G$
and $V_S$ are the gate and source voltages. The role of scattering of
current carriers in long channel transistors is modeled using the
mobility. For nanotransistors with ultra short channel lengths, there
are some deviations in the electrostatics from the long channel 
case~\cite{taur-book}. The role of scattering is however not well 
understood in nanotransistors. Most work
on nanotransistors use the drift diffusion equations which are 
applicable to long channel MOSFETs or fully ballistic calculations 
based on the Schroedinger equation. A detailed understanding of the 
influence of scattering is important as it is crucial in determining 
the on-current of nanotransistors. The role of scattering is however 
not straight forward to determine without a calculation
because scattering tends to change the carrier and current densities 
in the channel and extension regions, both spatially and energetically.
Further, the physics of this redistribution depends sensitively on the
channel and scattering lengths as demonstrated in this paper.

The aim of this paper is to model the exact influence of scattering at
different spatial locations along the channel and extension regions
of silicon n-MOSFETs. We consider only 
electron-phonon scattering, which is an important scattering mechanism
in devices with undoped channels. References~\cite{fischetti-jap-01a}
and \cite{fischetti-jap-01b} have recently pointed out that 
electron-electron and plasmon scattering may play an important role in
degrading nanotransistor characteristics. 
Electron-electron scattering in the drain side will lead to carriers 
having an energy larger than the source injection barrier. The 
resulting small tail of hot carriers~\cite{fischetti-iedm-95} will
be reflected back into the source-end, there by causing an increase in
the source injection barrier and a corresponding decrease in drain 
current. The modeling of these effects and interface roughness is beyond 
the scope of our current work. 

In our calculations, we consider the dual gate 
MOSFET~\cite{yan-apl-91,fischetti-iedm-92}, which is considered to
be a promising candidate for nanotransistors. The reason for this is
the large on-current and better scaling properties it offers, when 
compared to bulk-type 
MOSFETs~\cite{taur97,wong-iedm-97,pikus-apl-97,ren-iedm-00,chang-iedm-2000,walting-preprint}.

The outline of the paper is as follows. In section \ref{sect:results},
we present our simulation results on the role of scattering in 
nanotransistors, where we show that scattering is important throughout
the device, and not just in the source-end. This is followed by a
discussion explaining why drain-end scattering is important in 
nanotransistors (section \ref{sect:discussion}). In section 
\ref{sect:series-resistance}, we show that scattering in the extension
regions cannot be modeled as simple series resistances. We conclude in
section \ref{sect:conclusions}. All details of our method and 
approximations are given in the appendix.

\section{Where is scattering important?: Simulation Results}
\label{sect:results}

Two devices (Fig. \ref{fig:dg}) are simulated with the following 
parameters:

\noindent
\underline{Device A} (similar to the Purdue dual
gate MOSFET \cite{lundstrom-ieee-ted-02}.): 
Channel length ($L_{Ch}$) = 10 nm, channel extends from -5 nm to 5 nm,
channel thickness ($T_{Ch}$) = 1.5 nm, oxide thickness = 1.5 nm, gate
work function = 4.25 eV, doping in the extension regions = 1 E+20 
cm$^{-3}$, no doping in the channel, drain voltage ($V_D$) = gate 
voltage ($V_G$) = 0.6 V, and the dielectric constant of the oxide 
($\epsilon_{ox}$)=3.9.

\noindent
\underline{Device B}: Same as Device A, except that $L_{Ch} = 25$ nm, 
channel extends from - 12.5 nm to 12.5 nm and $V_G = 0.56$ V. In all
simulations involving this device, scattering is included only in
the channel.
\noindent
%\underline{Device C}: 
%Same as Device A, except that the $\epsilon_{ox}$= 20 and gate work 
%function = 4.3417 eV. Device C has a higher dielectric constant for the
%gate oxide and almost no drain induced barrier lowering (DIBL) when 
%compared to Device A. 
The gate length is equal to the channel length for both devices A and
B. The temperature assumed in all calculations in this paper is 300K.

We first discuss device A. To elucidate the role of scattering in 
different spatial regions, we calculate the drain current ($I_D$) as a
function of the right boundary of scattering, $Y_{R-Scatt}$. Scattering
is included from the edge of the source extension region (-20 nm) to 
$Y_{R-Scatt}$ in Fig. \ref{fig:IdvsYRscatt10}. The ballistic current is
1.92 mA\//$\mu$m, the value at $Y_{R-Scatt}=-20$ nm. The channel 
extends from -5 nm to +5 nm. The main points of this figure are:

(i) The decrease in current from the ballistic value due to scattering
in the source extension, channel and drain extension regions are 
11.5\%, 15.5\% and 4\% respectively. These values point to the well 
appreciated result that either reducing the length or flaring the 
source extension region will make a nanotransistor significantly more
ballistic.

(ii) The decrease in drain current due to scattering over the entire
channel is important. That is, scattering in the right half of the
channel (0 nm to 5nm) is almost as important as scattering in the left
half of the channel (-5 nm to 0 nm). 

(iii) The drain current continues to decrease significantly due to
scattering in the drain extension region. An important question is
if this decrease is simply a series resistance effect (see section
\ref{sect:series-resistance}).

We now present results for device B, whose channel length is two and a
half times larger than device A. The scattering times are nearly the
same for the two devices.  As a result of the larger channel length, 
the probability for a carrier to energetically relax is larger. Here,
we find that scattering in the left (-12.5 nm to 0 nm) and right (0 nm
to 12.5 nm) halves of the channel reduces the drain current by 32\% 
and 15\% respectively from the ballistic value, and the over all 
ballisticity (ratio of {\it Current with scattering} to {\it Ballistic
current}) is 53\% (dashed line of Fig. \ref{fig:IdvsYRscatt25}).
Again, this points to the importance of scattering in the drain-end.

In lieu of simulating devices with longer channel lengths, we increase
the scattering rate of device B. The scattering rate is increased by a
factor of five by artificially increasing the values of the deformation
potential quoted in reference \cite{lundstrom-book} by a factor of 
$\sqrt{5}$. Note that device B has almost no DIBL and that we 
self-consistently solve the Green's function and Poisson's equations
with the larger deformation potentials.
The ballisticity of device B with the larger scattering rate is 38\%,
and the current decreases by 60\% and 12\% of the ballistic value due
to scattering in the left and right halves of the channel respectively
(solid line of Fig. \ref{fig:IdvsYRscatt25}). It is also apparent from
Fig.  \ref{fig:IdvsYRscatt25} that the effect of scattering on drain 
current becomes relatively smaller as $Y_{R-Scatt}$ approaches the 
drain-end (12.5 nm).

\section{Discussion}
\label{sect:discussion}

The results of section \ref{sect:results} show that scattering 
{\it at all} locations in the channel is 
important in determining the drain current of nanoscale MOSFETs. We 
first discuss device A.  For device A, the scattering time 
($\hbar/2|Im\left(\Sigma^r_{phonon}\right)|$) at an energy of $E_b + 
26$ meV is 50 fs and 24 fs in the source
and drain-ends respectively. The scattering times are comparable to 
the semiclassical transit time of 26 fs (Table 
\ref{table:time}). The scattering (11 nm)  and channel lengths (10 nm)
are hence comparable (Table \ref{table:time}). It is interesting to 
note that for this device, the argument that the energetic 
redistribution of electrons in the channel to states with kinetic 
energy in the transport direction well below $E_b$ will make drain-end
scattering ineffective fails. 

To understand why drain-end scattering is important for the parameters
in device A, it is useful to plot the change in barrier height ($E_b$)
with $Y_{R-Scatt}$. Fig.  \ref{fig:IdvsYRscatt10} shows $E_b$ as a 
function of $Y_{R-Scatt}$. It is noted that $E_b$ first decreases
and then increases, with increase in $Y_{R-Scatt}$. The decrease of 
$E_b$ for $-20\mbox{ nm} < Y_{R-Scatt} < -4 \mbox{ nm}$ is due to the
potential drop in the source extension region arising from the 
increasing series resistance. Note that the location of the source 
injection barrier ($Y_b$) is  -4 nm (Fig. \ref{fig:pot_profile}). For
$Y_{R-Scatt} > Y_b$, $E_b$ increases with $Y_{R-Scatt}$. The reason for
the increase in $E_b$ are the electrons reflected towards the source 
from the right of $Y_b$. Electrostatics, more or less demands that the
charge in the gate should be approximately $C_{ox} 
(V_G-V_S)$~\cite{natori-jap-94,lundstrom-edl-97}, like in long channel
MOSFETs~\cite{taur-book}. So, $E_b$ floats to higher energies to 
compensate for the increase in electron density from the reflected 
electrons. This increase in $E_b$  contributes significantly to the
decrease in the drain current even due to scattering in the right half
of channel (0 nm to 5 nm). The increase in $E_b$ with increase in 
$Y_{R-Scatt}$ becomes smaller in the right end of Fig.
\ref{fig:IdvsYRscatt10} because the electrons scattered here contribute
less significantly to the channel charge, as will be apparent from the
discussion below.

We now discuss device B. Device B is different from device A in that 
its channel length is two and a half times longer than that of device
A. The importance of scattering in the right half of the channel is 
obvious for device B from the dashed line of Fig. 
\ref{fig:IdvsYRscatt25}. Here, scattering in the left (-12.5 nm to 0 
nm) and right (0 nm to 12.5 nm) halves of the channel reduce the drain
current by 32\% and 15\% respectively from the ballistic value.
To complement the discussion of device A in terms of $E_b$, we will 
discuss device B in terms of another useful quantity: 
$J(Y,E)$, which is the current 
distribution as a function of total energy E at $Y$. $J(Y,E)$ gives
us partial information about the energetic redistribution of current 
due to scattering  (see end of current section). When the channel 
length is comparable to the scattering length, $J(Y,E)$ is peaked in 
energy above $E_b$, in the right half of the channel 
(Fig. \ref{fig:paper_CURvsEvsY} (a)). Scattering causes reflection 
of this current towards the source. This is the first reason for the 
reduction in drain current.  The second reason is that the reflected
electrons lead to an increase in the channel electron 
density (classical MOSFET electrostatics). As the charge in the channel
should be approximately $C_{ox} (V_G-V_S)$, the source injection 
barrier $E_b$ floats to higher energies to compensate for the reflected
electrons. The increase in $E_b$ leads to a 
further decrease in drain current due to scattering in the right half
of the channel.

To gain further insight into the role of carrier relaxation, we now
discuss device B when the scattering length is five times smaller.
The scattering length $L_{scatt}$ is defined in Table \ref{table:time}.
Scattering in the right half of the channel for $L_{scatt}=2.2$ nm is
significantly less detrimental to the drain current relative to 
scattering in the left half of the channel, when compared to the device
with $L_{scatt}=11$ nm. As $L_{Ch}$ (25 nm) is much larger than 
$L_{scatt}$ (2.2 nm),
multiple scattering events now lead to an energy distribution of 
current that is peaked well below the source injection barrier in the
right half of the channel as shown in Fig. \ref{fig:paper_CURvsEvsY}
(b). The first moment of energy (mean) with respect to the current 
distribution function, which is defined by 
$\frac{\int dE E J(Y,E)}{\int dE J(Y,E)}$, is also shown in 
Fig. \ref{fig:paper_CURvsEvsY}.
This mean also shows that the carriers relax in a manner akin to bulk
MOSFETs as a function of Y in Fig. \ref{fig:paper_CURvsEvsY} (b).
Carriers reflected in the right half of the channel can no longer reach
$Y_b$ due to the large barrier to the left, and so contribute less 
significantly to the charge density. Thus, explaining the diminished 
influence of scattering in the right half of the channel relative to 
the left half of the channel, for devices with the channel length much
larger than the scattering length.

The above discussion would be incomplete without discussing the
electrostatic potential profiles, with and without scattering. The 
solid line in Fig. \ref{fig:paper_POTvsY} is the electrostatic 
potential in the ballistic limit. Increasing $Y_{R-Scatt}$ from - 2.5
nm to 2.5 nm causes $E_b$ to increase because of carriers reflected 
towards the source. Further increase in $Y_{R-Scatt}$ to 7.5 nm causes
very little increase in $E_b$ because scattering in the right half
of the channel is less effective in changing the channel electron
density. The electrostatic potential changes appreciably to the right
of $Y_b$ due to scattering. It is also interesting to note that the 
electrostatic 
potential drop for $Y_{R-Scatt}=7.5$ nm is linear to the right of $Y_b$
compared to the ballistic case because of scattering in the channel.

We now comment briefly on two issues: 

\noindent
-  The quantity $E_b - 2kT$ that has been discussed before in 
references \cite{lundstrom-ieee-ted-02} and \cite{price-79}. 

\noindent
- The influence of elastic scattering without any inelastic
scattering. 

For devices A and B, the potential profile in the right half of the 
channel is well below $E_b - 2kT$. Yet, scattering in the right half of
the channel is detrimental to drain current, relative to scattering in
the left half of the channel. The reason for this are the hot electrons
in the right half of the channel that are reflected to the source-end
/ $Y_b$. However, if the scattering rate in the left
half of the channel is large enough to energetically relax the 
electrons to energies comparable to $E_b - 2kT$, then the scattering of
these electrons in the drain-end are relatively less detrimental
to the reduction in drain current because the carriers cannot 
easily gain an energy of few times the thermal energy. This phenomenon
of the diminished role of scattering in the channel at the drain-end
relative to the source-end because of thermalized carriers is seen in
Fig. \ref{fig:IdvsYRscatt25} (solid line).

In the presence of elastic scattering processes such as interface 
roughness scattering, the electron does not loose total energy. 
However, the kinetic energy in the transport direction can diminish at
the expense of a corresponding gain in $\frac{\hbar^2 k_z^2}{2m_z^n}$.
The additional density of states for scattering that is available
in the drain-end in comparison to the source-end will also make
drain-end scattering less effective than source-end scattering.
While we included such process in our calculations, the quantity
$J(Y,E)$ captures only the effect of
change in total energy. A physically motivated study quantifying the
relative roles of elastic and inelastic scattering will be a useful
future study.

\section{Failure of the classic series resistance picture for 
nanotransistors}
\label{sect:series-resistance}

We ask the question if scattering in the extension regions is a simple
series resistance. The classic series resistance 
picture~\cite{taur-book} relates the current in a device with long 
extension regions to the current in the same device without (or with 
much smaller) extension regions. The relationship is particularly 
simple for the case where the series resistance in the source extension
region is negligible~\cite{taur-book},
\begin{eqnarray}
I_D^{scatt}(V_D) \sim  I_D^{no scatt}(V_D - \delta V_D) \mbox{ ,}
                                                     \label{eq:series}
\end{eqnarray}
where, $I_D^{scatt} (V_D)$ and $I_D^{no scatt} (V_D - \delta V_D) R_D)$
are the drain currents with and without scattering in the drain 
extension region, at drain biases of $V_D$ and $V_D - \delta V_D$ 
respectively. $\delta V_D=I_D^{scatt} (V_D) R_D$, is the electrostatic
potential drop in the drain extension region, which has a series
resistance of $R_D$.  To answer the question on the 
appropriateness of the classic series resistance picture, we consider
a case where the channel and source extension region are ballistic. 
Scattering is introduced only in the drain extension region with
deformation potentials that are $\sqrt{5}$ times larger than in 
silicon (scattering time is 5 times smaller). 

Fig. \ref{fig:paper-series} shows the decrease in drain current with
$Y_{R-Scatt}$. The striking point of Fig. \ref{fig:paper-series} is
the super-linear decrease of drain current. The $I_D(V_D)$ curves 
(inset of Fig. \ref{fig:paper-series}) predict a significantly smaller
decrease in drain current with increase in $Y_{R-Scatt}$ when Eq. 
\ref{eq:series} is used. It is helpful to estimate the drain current 
from Eq. \ref{eq:series} and compare it to the calculated value. For 
Device A in Fig. \ref{fig:paper-series}, the voltage drop in the drain
extension region with scattering is approximately 100 mV (plot not 
shown). Now, if Eq. \ref{eq:series} is used to estimate the drain 
current with scattering in the drain extension region and if we take 
$\delta V_D= 200$ mV, which is larger than the estimated 100 mV, then we
find the drain current to be 1.83 mA/$\mu$m (inset of Fig. 
\ref{fig:paper-series}).  The calculated drain current is however much
 lower at 1.38 mA/$\mu$m!

%We have also considered a device with negligible DIBL (device C, see
%inset of Fig.  \ref{fig:paper-series}). device C illustrates that 
%the effect discussed exists even in devices without DIBL, and is
%different from the classical series resistance.

The physics of the large reduction in drain current for the smaller
values of $Y_{R-Scatt}$ is essentially that discussed in section
\ref{sect:discussion}: When scattering in the channel does not 
effectively thermalize carriers, the current distribution is peaked at
energies above $E_b$, upon carriers
exiting the channel. Scattering in the drain extension region then 
causes reflection of electrons towards the source-end. As a result, 
$E_b$ increases so as to keep the electron density in the channel
approximately $C_{ox} (V_G-V_S)$. The drain current decreases 
dramatically as a result of the increase in $E_b$. Admittedly, this 
argument in terms of $C_{ox} (V_G-V_S)$ is over simplified but it seems 
to capture the essential point. The main point is that
if carriers are not relaxed upon exiting the channel (as would be the
case for nano-transistors), then, the drain extension region cannot be
modeled
by a simple series resistance. That is, Eq. (\ref{eq:series}) fails for 
nano-transistors where the channel length is comparable to the
scattering length. The effect of the drain extension region in causing
a reduction in drain current would be small in the following cases:

{\it (i)} The channel is much longer than the scattering length such 
that the carriers exiting the channel at the drain-end are 
energetically relaxed / thermalized. Then, the modeling of the drain 
extension region as a simple series resistance would be appropriate. 
This is seen in the right end of Fig. \ref{fig:paper-series}, where, 
upon sufficient relaxation of electrons, the decrease in current with
increase in $Y_{R-Scatt}$ becomes much smaller.

{\it (ii)} The drain extension region rapidly flares out. Then, the 
probability for a scattered electron to return to the source-end will 
be small due to the larger number of modes available in the drain 
extension region. A careful analysis on how fast the drain extension 
region flares out should also take into account the role of the Miller
effect.

\section{Conclusions}
\label{sect:conclusions}

\begin{figure}[htbp]
  \begin{center}
    \leavevmode
     \epsfxsize=5.375in
    \epsfbox{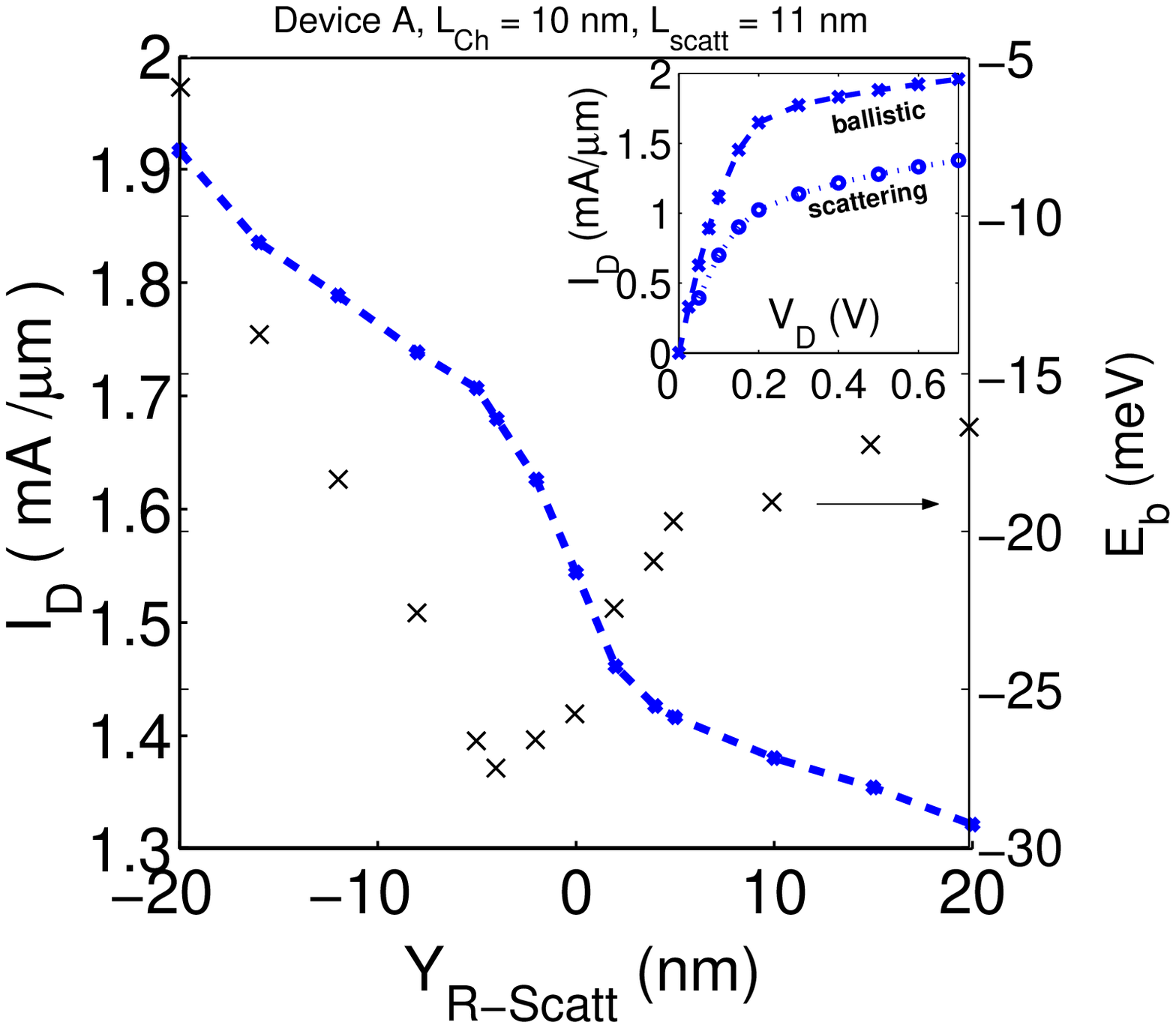}
  \end{center}
\vspace{2in}
\caption{\label{fig:IdvsYRscatt10}}
\end{figure}

\begin{figure}[htbp]
\begin{center}
     \leavevmode
     \epsfxsize=5.375in
    \epsfbox{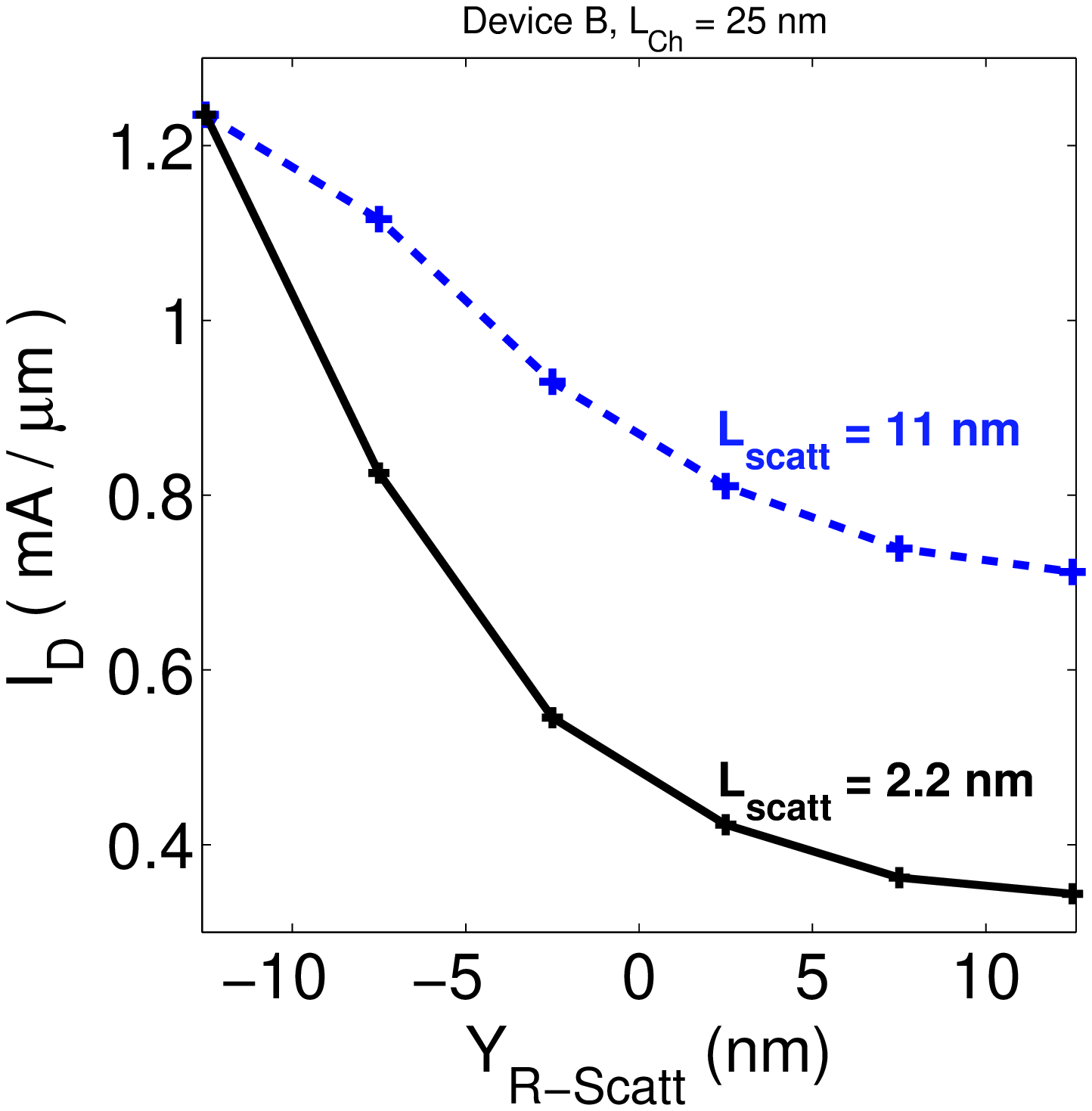}
  \end{center}
\vspace{2in}
\caption{\label{fig:IdvsYRscatt25}}
\end{figure}

In conclusion, we find that the potential profile, channel and 
scattering length scales play an important role in determining the 
{\it relative} importance of scattering at different locations along 
the channel of a nanotransistor. In devices where the channel length
is comparable to the scattering length, the role of scattering in the 
drain-end (right half of the channel) is comparable to the role of 
scattering in the source-end (left half of the channel), in reducing 
the drain current (Fig. \ref{fig:IdvsYRscatt10} and dashed line of 
Fig. \ref{fig:IdvsYRscatt25}). This is contrary to a belief that
scattering is significantly more important in the source-end of the 
device. The reason for the detrimental role of scattering in the
drain-end are the hot carriers in the drain-end.
When the channel length is much larger than the scattering 
length, then scattering in the source-end becomes relatively more 
important than scattering in the drain-end (solid line of Fig. 
\ref{fig:IdvsYRscatt25}). In this case, we stress that it is the
energetic redistribution of carriers due to scattering in the 
source-end to energies below the source injection barrier ($E_b$) that
makes scattering in the drain-end relatively less detrimental to the 
drain current. 

The classical series resistance picture for modeling the narrow
extension regions fail for nanotransistors. The reason for this
failure are the hot carriers entering the drain extension region.
A straight forward option to enable the usage of the series resistance
picture is to push the region treated as a drain series resistance 
further to the right, such that all carriers entering this region
are energetically relaxed. A more interesting option of altering
the classical series resistance picture to account for the hot carriers
in the drain end of nanotransistors was not considered in this paper.

The relative importance of scattering in the drain-end of 
nanotransisors, where the channel length is comparable or smaller 
than the scattering length, points to the importance of making the 
extension regions small. Long extension regions in nanotransistors will
affect the performance (drive current) much more adversely than in 
long channel transistors.

\appendix
\label{sec:approach}

\begin{figure}[htbp]
  \begin{center}
    \leavevmode
     \epsfxsize=5.375in
    \epsfbox{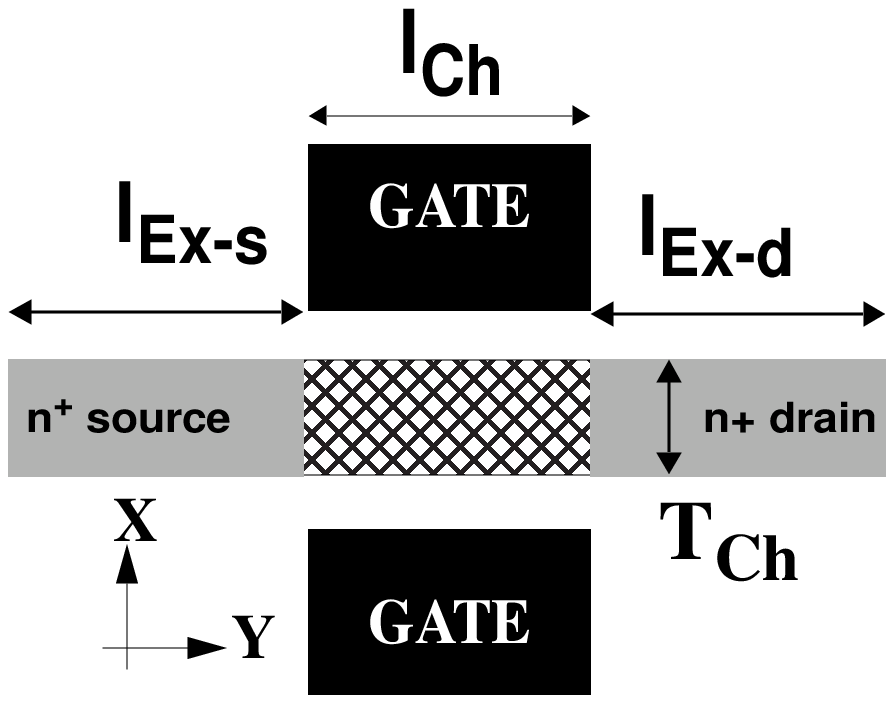}
  \end{center}
\vspace{2in}
\caption{\label{fig:dg}}
\end{figure}

The approach consists of solving the nonequilibrium Green's function
and Poisson's equations. The effective mass Hamiltonian considered is,
\begin{eqnarray}
H = \sum_b -\frac{\hbar^2}{2} \left[\frac{d}{dx} \left(\frac{1}{m^b_x}
\frac{d}{dx}\right) + \frac{d}{dy} \left(\frac{1}{m^b_y}\frac{d}{dy}
\right) + \frac{d}{dz} \left( \frac{1}{m^b_z}\frac{d}{dz} \right) \right]
+ V(x,y) \mbox{,}
\end{eqnarray}
where $(m^b_x,m^b_y,m^b_z)$ are the (x, y, z) components of the 
effective mass in valley $b$ of silicon, and the potential does not 
vary in the $z$ direction. The gate oxides are treated as hard walls,
the channel is extremely narrow (1.5 nm), the drain and gate biases are
smaller than 0.7 V, and the dual gate FET is perfectly symmetric in the 
$X$-direction of Fig. \ref{fig:dg}. 
The first three subband energy levels in the source
extension region are approximately equal to 173 meV, 691 meV (both due
to $m_y=0.98 m_0$) and 891 meV (due to $m_y=0.19 m_0$) above the bulk
conduction band. The Fermi energy of bulk silicon at the doping
density considered (1E+20 cm$^{-3}$) is approximately 60 meV above the
conduction band. For the doping density considered, electrons
are primarily injected from the source into the first subband. At
the drain end, more than one energy level can in principle contribute
to current. As only a few subbands are populated, we model transport
in these subbands in an approximate way using the 1D Schroedinger
equation as outlined below.
We find the spatially dependent subband energies
$E_n(y)$ by solving Schroedinger's equation at each $y$-cross section
($y$ is only a parameter),
\begin{eqnarray}
\left[-\frac{\hbar^2}{2 m_x^b} \frac{d}{dx}
\left(\frac{1}{m^b_x}\frac{d}{dx} \right) + V(x,y) \right] \;
\Psi_n(x,y) = E_n (y) \Psi_n(x,y) \mbox{ .} \label{eq:Sch}
\end{eqnarray}
$n={\nu,b}$, where $\nu$ and $b$ represent the quantum number due
to quantization in the X-direction and the valley respectively.
The valley indices $b$ are required in the calculations of the 
self-energies for scattering as will be discussed below. In our
calculation, we typically retain only the three lowest energy levels.
Coupling between the
subbands is neglected except via phonon coupling. For the device
dimensions and voltages considered, reference \cite{venugopal-jap-2002}
found the approximation of considering decoupled subbands to hold good
for ultra thin body phase coherent MOSFETs. We solve the following 
equations for the Green's functions,
\begin{eqnarray}
&& \;\;\;\;\;\;\;\;\;\;\;\;\;\;\;\;
 \left[ E - \frac{\hbar^2 k_z^2}{2m_z^n} - \left(-\frac{\hbar^2}{2}
\frac{d}{dy} \left(\frac{1}{m^n_y}\frac{d}{dy} \right) + E_n(y) \right)
\right] G^r_{n} (y,y^\prime,k_z,E)  \nonumber \\
&& 
- \int dy_1 \; \Sigma^r_{n} (y,y_1,k_z,E) G^r_{n} (y_1,y^\prime,k_z,E) =
                 \delta(y-y^\prime)  \mbox{ , and} \label{eq:Gr_scatt}
\end{eqnarray}
\begin{eqnarray}
&&\!\!\!\!\!\!\!\!\!\!\!\!\!\!\!\!
 \left[ E - \frac{\hbar^2 k_z^2}{2m_z^n} - \left(-\frac{\hbar^2}{2}
\frac{d}{dy} \left(\frac{1}{m^n_y}\frac{d}{dy} \right) + E_n(y) \right)
\right] G^\alpha_{n} (y,y^\prime,k_z,E)  \nonumber \\
&& 
- \int dy_1 \; \Sigma^r_{n} (y,y_1,k_z,E)
G^\alpha_{n} (y_1,y^\prime,k_z,E) =
\int dy \; \Sigma^\alpha_{n} (y,y_1,k_z,E)
           G^a_{n} (y_1,y^\prime,k_z,E) \mbox{ ,} \label{eq:Gl_scatt}
\end{eqnarray}
where, $\alpha \in >,<$. $m_y^n$ and $m_z^n$ are the effective masses
of silicon in the $y$ and $z$ directions that give rise to subband 
index $n$.

The self-energies, $\Sigma^{r, >, <}_{n}$ can be written as,
\begin{eqnarray}
\Sigma^\alpha_{n} &=& \Sigma^\alpha_{n,C} + \Sigma^\alpha_{n,Phonon}
\mbox{, where} \\
\Sigma^\alpha_{n,Phonon} &=& \Sigma^\alpha_{n,el} + 
                                              \Sigma^\alpha_{n,inel}
\mbox{ .}
\end{eqnarray}
$\Sigma^\alpha_{n,C}$ is the self-energy due to the leads. The phonon
self-energy $\Sigma^\alpha_{n,Phonon}$ consists of two terms,
$\Sigma^\alpha_{n,el}$ due to elastic and $\Sigma^\alpha_{n,inel}$ due
to inelastic scattering. The self-energy due to the leads is non zero
only at the first (source) and last (drain) grid points because gate 
tunneling is neglected.

The following common approximations to calculate the phonon
self-energies are used:
(i) Phonon scattering is treated only within the self-consistent Born
    approximation,
(ii) The phonon bath is assumed to always be in equilibrium, and so
their occupation numbers are given by the Bose-Einstein distribution
function with a spatially independent temperature.
(iii) The correlation between subbands $n$ and $n^\prime$
($\neq n$) are neglected.
(iv) Scattering due to phonons is assumed to be isotropic. That is,
the scattering rate from $(k_z,E)$ to $(k_z^\prime,E^\prime)$ does
not depend on $k_z$ and $k_z^\prime$. This approximation is 
computationally advantageous because the self-energies due to phonon 
scattering appear only as diagonal terms in Eqs. \ref{eq:Gr_scatt}
and \ref{eq:Gl_scatt}.
One can derive from these assumptions that the self-energies due to
electron-phonon scattering at grid point $y_i$ are given
by \cite{mahan-physrep-87,lake-jap-97},
\begin{eqnarray}
\Sigma^\alpha_{el,n} (y_i,E) &=& \sum_{n^\prime}
D_{n,n^\prime}^{el}
\frac{\sqrt{m_z^{n^\prime}}}{\pi \hbar \sqrt{2}} \int dE_z
\frac{1}{\sqrt{E_z}} G^\alpha_{n^\prime}(y_i,E_z,E) \mbox{ ,}
\label{eq:el_self_en_r}
\end{eqnarray}
\begin{eqnarray}
&&\Sigma^<_{inel,n} (y_i,E) = \sum_{n^\prime,\eta}
D_{n,n^\prime}^{i,\eta} \frac{\sqrt{m_{z}^{n^\prime}}}{\pi \hbar \sqrt{2}}
\int dE_z \frac{1}{\sqrt{E_z}}  \nonumber \\
&& \;\;\;\;\;\;\left[n_B(\hbar \omega_\eta)
G^<_{n^\prime}(y_i,E_z,E-\hbar \omega_\eta)  +
(n_B(\hbar \omega_\eta)+1)
G^>_{n^\prime}(y_i,E_z,E+\hbar \omega_\eta) \right] \mbox{ ,}
\label{eq:inel_self_en_<} 
\end{eqnarray}
and
\begin{eqnarray}
&&\Sigma^>_{inel,n} (y_i,E) = \sum_{n^\prime,\eta}
D_{n,n^\prime}^{i,\eta} \frac{\sqrt{m_{z}^{n^\prime}}}{\pi \hbar \sqrt{2}}
\int dE_z \frac{1}{\sqrt{E_z}}  \nonumber \\
&& \;\;\;\;\;\;\left[n_B(\hbar \omega_\eta)
G^>_{n^\prime}(y_i,E_z,E+\hbar \omega_\eta)  +     
(n_B(\hbar \omega_\eta)+1)
G^>_{n^\prime}(y_i,E_z,E-\hbar \omega_\eta) \right] \mbox{ .}
\label{eq:inel_self_en_>} 
\end{eqnarray}
$\alpha \in >,<,r$ in Eq. \ref{eq:el_self_en_r}, $\eta$ represents
the phonon modes, and the square of the matrix elements for phonon 
scattering are given by,
\begin{eqnarray}
D_{n,n^\prime}^{el} &=& (\delta_{\nu,\nu^\prime} + \frac{1}{2})  
\delta_{b,b^\prime} \frac{D_A^2 kT}{\rho v^2} \\
D_{n,n^\prime}^{i,\eta} &=& (\delta_{\nu,\nu^\prime} + \frac{1}{2})
\left[\delta_{b,b^\prime} \frac{D_{g\eta}^2 \hbar}{2\rho \omega_{g\eta}}
+ (1-\delta_{b,b^\prime}) \frac{D_{f\eta}^2 \hbar}{\rho \omega_{f\eta}}
\right]   
\end{eqnarray}
The contribution to elastic scattering is only from acoustic phonon
scattering. The values of the deformation potential, $D_A$, $D_{g\eta}$
and $D_{f\eta}$, and phonon frequencies $\omega_{g\eta}$ and 
$\omega_{f\eta}$ are taken from \cite{lundstrom-book}. $\rho$ is the 
mass density, $k$ is the Boltzmann constant, $T$ is the temperature and
$v$ is the velocity of sound. $b$ and $b^\prime$ are indices 
representing the valley. The following scattering processes are 
included:
acoustic phonon scattering in the elastic approximation and g-type
intervalley scattering with phonon energies of 12, 19 and 62 meV.
It was verified that f-type (19, 47 and 59 meV phonon) intervalley
scattering did not significantly change our results and conclusions.
This can be rationalized by noting that f-type intervalley scattering
involves subbands with energies higher than the lowest subband. 
In the regions, where scattering was not included, the deformation
potential was set equal to zero.

$\Sigma^r_{inel,n}$ can be obtained using the Kramers-Kronig
relationship,
\begin{eqnarray}
Re\left[\Sigma^r_{inel,n} (y_i,E)\right] &=& 
\frac{1}{\pi} \;\; \mbox{P } \int dE^\prime
\frac{Im \left[\Sigma^r_{inel,n} (y_i,E^\prime)\right]}{E^\prime-E}
\mbox{ and } \\
Im\left[\Sigma^r_{inel,n} (y_i,E)\right] &=& \frac{1}{2i} \left[
\Sigma^>_{inel,n} (y_i,E) - \Sigma^<_{inel,n} (y_i,E)
\right] \mbox{ ,} 
\end{eqnarray}
where $P$ stands for the principal part of the integral. Note that the
self-energies due to electron-phonon scattering depend only on the 
total energy $E$ (and not on $k_z$) due to the assumption of isotropic
scattering.

The self-energy due to
phonon scattering, has real and imaginary parts, both of which vary
with energy. The imaginary part of the electron-phonon self-energy
which is central to our calculations is responsible for scattering
induced broadening of energy levels and energetic redistribution of
carriers. The real part of the self-energy which contributes to the 
shift of the quasi-particle energy levels, appears as a real potential
(like the electrostatic potential) in the Green's function equations
(Eqs. \ref{eq:Gr_scatt} and \ref{eq:Gl_scatt}).
To evaluate the importance of the real part of the self-energy in our
calculations, we performed simulations with acoustic phonon scattering
in silicon, with and without the real part of the self-energy included.
We find that the drive current calculated with the real part of the
self-energy set to zero in general agrees to within 2 
percent of the current calculated with the real part of the self-energy
included. This result is not totally surprising because MOSFET 
electrostatics tends to shift the potential profile appropriately to 
determine the correct charge under the gate. In the calculations 
presented in this paper, the real part of the self-energy is set to 
zero.

In the numerical solution, we consider $N$ uniformly spaced grid 
points in the $Y$-direction with the grid spacing equal to $\Delta y$.
The discretized form of Eqs. \ref{eq:Gr_scatt} and \ref{eq:Gl_scatt}
are:
\begin{eqnarray}
 A_{i,i}   G^r_{n} (y_i,y_i^\prime,k_z,E)
 + A_{i,i+1} G^r_{n} (y_{i+1},y_i^\prime,k_z,E)
 + A_{i,i-1} G^r_{n} (y_{i-1},y_i^\prime,k_z,E) =
\frac{\delta_{i,i^\prime}}{\Delta y}
              \mbox{ , and} \label{eq:Gr-scatt-discrete}
\end{eqnarray}
\begin{eqnarray}
&& \!\!\!\!\!\!\!\!\!\!\!\!\!\!\!\!\!\!\!\!\!\!\!\!\!\!\!\!\!\!\!\!
A_{i,i}   G^\alpha_{n} (y_i,y_i^\prime,k_z,E)
 + A_{i,i+1} G^\alpha_{n} (y_{i+1},y_i^\prime,k_z,E)
 + A_{i,i-1} G^\alpha_{n} (y_{i-1},y_i^\prime,k_z,E) = \nonumber \\
&& \;\;\;\;\;\;\;\;\;\;\;\;\;\;\;\;\;\;\;\;\;\;\;\;
 \;\;\;\;\;\;\;\;\;\;\;\;\;\;\;\;\;\;\;\;\;\;\;\;
 \;\;\;\;\;\;\;\;\;\;\;\;\;\;\;\;\;\;\;\;\;\;\;\;
\Sigma_n^\alpha (y_i,E) G^a_{n} (y_i,y_i^\prime,k_z,E)
              \mbox{ , } \label{eq:G<-scatt-discrete}
\end{eqnarray}
where,
\begin{eqnarray}
A_{i,i} &=& E - \frac{\hbar^2 k_z^2}{2m_z^n} -
                \frac{\hbar^2}{m^n_y \Delta y^2} -
                 E_n(y_i) - \Sigma^r_n (y_i,k_z,E) \mbox{ and }\\
A_{i\pm1,i} &=& + \frac{\hbar^2}{2m_z^n \Delta y^2}
\end{eqnarray}
The self-energy due to the source and drain leads contribute only to
grid point 1 (left end of the source extension region) and grid point 
N (right end of the drain extension region), and are given
by \cite{datta-book}:
$\Sigma^r_{n,C}(y_1,k_z,E) = (\frac{\hbar^2}{2m_z^n \Delta y^2})^2
g_s(k_z,E)$,
$\Sigma^r_{n,C}(y_N,k_z,E) = (\frac{\hbar^2}{2m_z^n \Delta y^2})^2
g_d(k_z,E)$,
$\Sigma^<_{n,C}(y_1,k_z,E) = -2i \mbox{Im}(\Sigma^r_{n,C}(y_1,k_z,E))
f_{s}(E)$, $\Sigma^<_{n,C}(y_N,k_z,E) =$

\noindent
$-2i \mbox{Im}(\Sigma^r_{n,C}(y_N,k_z,E))
f_{d}(E)$,
$\Sigma^>_{n,C}(y_1,k_z,E) =  2i \mbox{Im}(\Sigma^r_{n,C}(y_1,k_z,E))
[1-f_{s}(E)]$, and

\noindent
$\Sigma^>_{n,C}(y_N,k_z,E) = 2i \mbox{Im}(\Sigma^r_{n,C}(y_N,k_z,E))
[1-f_{d}(E)]$, where $y_1$ an $y_N$ are the left (source-end) and right
(drain-end) most grid points respectively, $g_s(k_z,E)$ and
$g_d(k_z,E)$ are the surface Green's functions of the source and drain
leads respectively, and $f_{s}$ and $f_{d}$ are the Fermi functions in
the source and drain contacts respectively.

The non equilibrium electron and current densities are calculated in
both the channel and extension regions using the algorithm for $G^<$ in
\cite{svizhenko-jap-02}, which avoids full inversion of the $A$ matrix.
For completeness, we state the expressions for the electron and current
densities used \cite{svizhenko-jap-02},
\begin{eqnarray}
n_n(y_i,k_z,E) &=& - iG_n^<(y_i,y_i,k_z,E) \label{eq:dens}  \\
J_n(y_i,k_z,E) &=& \frac{e}{\hbar} \sum_n 
                 \frac{\hbar^2}{2m_y^n \Delta y^2} 
      [ G_n^<(y_i,y_{i+1},k_z,E) - G_n^<(y_{i+1},y_i,k_z,E)] \mbox{ .}
   \label{eq:current}
\end{eqnarray}  
Note that Eqs. \ref{eq:dens} and \ref{eq:current} do not include
spin and valley degenaracies. The total electron and current densities
at grid point $y_i$ are given by,
\begin{eqnarray}
n(y_i) &=& 2 \sum_n \frac{\sqrt{m_z^{n^\prime}}}{\pi \hbar \sqrt{2}}
\int \frac{dE}{2\pi} \int dE_z \frac{1}{\sqrt{E_z}} n_n(y_i,E_z,E) \\
J(y_i) &=& 2 \sum_n \frac{\sqrt{m_z^{n^\prime}}}{\pi \hbar \sqrt{2}}
\int \frac{dE}{2\pi} \int dE_z \frac{1}{\sqrt{E_z}} 
J_n(y_i,E_z,E) \mbox{ ,}
\end{eqnarray}
where the prefactor of 2 in the above equations account for two
fold spin degenaracy. While the transport equations are 
solved in one dimension, we solve Poisson's equation in two dimensions.
The two dimensional electron density used in Poisson's equation is 
computed from Eqs. (\ref{eq:Sch}) and (\ref{eq:dens}) using,
\begin{eqnarray}
n(x_i,y_i,k_z,E) = n_n(y_i,k_z,E) |\Psi_n(x_i,y_i)|^2 \mbox{ .}
\end{eqnarray} 
The boundary conditions to Poisson's and Green's function equations are
applied at the ends of the source and drain extension regions (left and
right ends of the source and drain extension regions shown in Fig.
\ref{fig:dg}).
In solving the the Green's function and Poisson's equation, note that
an applied bias corresponds to a difference in the Fermi levels used
in the source and drain regions. The electrostatic potential at the
left and right most grid points of the source and drain extension
regions respectively are calculated self consistently using the
boundary conditions. 

Finally, we make a comment on the need for solving quantum mechanical
equations
to capture the essential effect of hot carriers, described in this
paper. The phase of the electron is not central to the physics
described in our paper (though the exact value of the drain current
depends on it). In calculating the drain current, the quantum
mechanical effects of quantization in the X-direction and tunneling
along the Y-direction (Fig. \ref{fig:dg}) can be accounted for
semiclassically. So, we feel that a method such as the Monte Carlo
method approach to nanotransistors \cite{fischetti-prb-93}, which 
keeps track of the details of the energetic redistribution of electrons
at various spatial locations, will well describe many aspects of the 
role of scattering.

\bibliography{scattbib} \bibliographystyle{unsrt}

%\begin{thebibliography}{100}

%\end{thebibliography}{100}
\clearpage

{\bf Figure Captions:}

\vspace{0.5in}

{\bf Fig. \ref{fig:dg}:}
Schematic of a Dual Gate MOSFET (DG MOSFET). Ex-s and Ex-d are the
extension regions and the hatched region is the 
channel. The white region between the source / drain / channel and 
gate is the oxide. The device dimension normal to the 
page is infinite in extent.

\vspace{0.5in}

{\bf Fig. \ref{fig:IdvsYRscatt10}:}
Plot of drain current ($I_D$) versus the right boundary of scattering
($Y_{R-Scatt}$) for device A. The scattering time is comparable to the
transit time through the channel. Scattering is included from -20 nm to 
$Y_{R-Scatt}$. Note that scattering in the right half of the channel 
(0 nm to 5 nm), which is to the right of the '$k_BT$ layer',
is almost as deleterious to current flow as scattering in the left
half of the channel (-5 nm to 0nm). The black crosses represent $E_b$
as a function of $Y_{R-Scatt}$.
Inset: Ballistic $I_D$ versus $V_D$ for $V_G=$ 0.6 V, showing 
substantial DIBL. Scattering is included both in the channel and
extension regions.

\vspace{0.5in}

\begin{figure}[htbp]
  \begin{center}
    \leavevmode
     \epsfxsize=5.375in
    \epsfbox{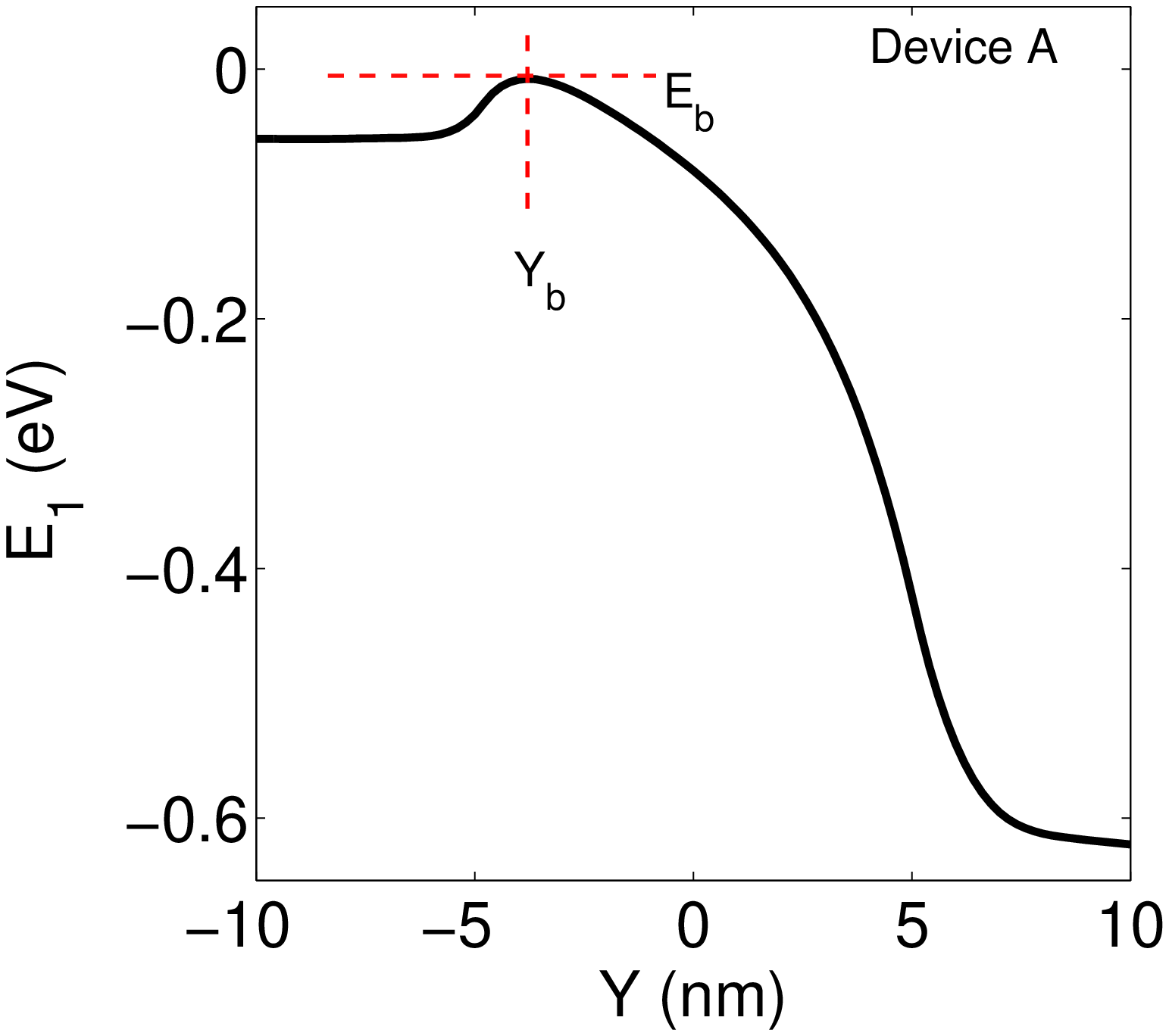}
  \end{center}
\vspace{2in}
\caption{\label{fig:pot_profile}}
\end{figure}

{\bf Fig. \ref{fig:pot_profile}:}
Energy of the lowest subband ($E_1$) versus $Y$ for device A in the
ballistic limit. $E_b$ and $Y_b$ are the energy and position of the
source injection barrier respectively. Potential = $-\frac{E_1}{e}$.

\vspace{0.5in}

{\bf Fig. \ref{fig:IdvsYRscatt25}:}
Plot of drain current versus $Y_{R-Scatt}$ for device B. Scattering is
included from -12.5 nm to $Y_{R-Scatt}$. For $L_{scatt}=11$ nm (dashed
line) and $2.2$ nm (solid line), the effect of scattering in the right
half of the channel (0 nm to 12.5 nm) corresponds to nearly a third and
sixth respectively of the total reduction in drain current. This figure
points to the relatively smaller role of drain-end scattering in 
comparison to source-end scattering, when $L_{Ch}$ becomes much larger
than $L_{scatt}$.  Scattering is included only in the channel for both
cases.

\vspace{0.5in}

\begin{figure}[htbp]
\begin{center}
     \leavevmode
     \epsfxsize=3.375in
    \epsfbox{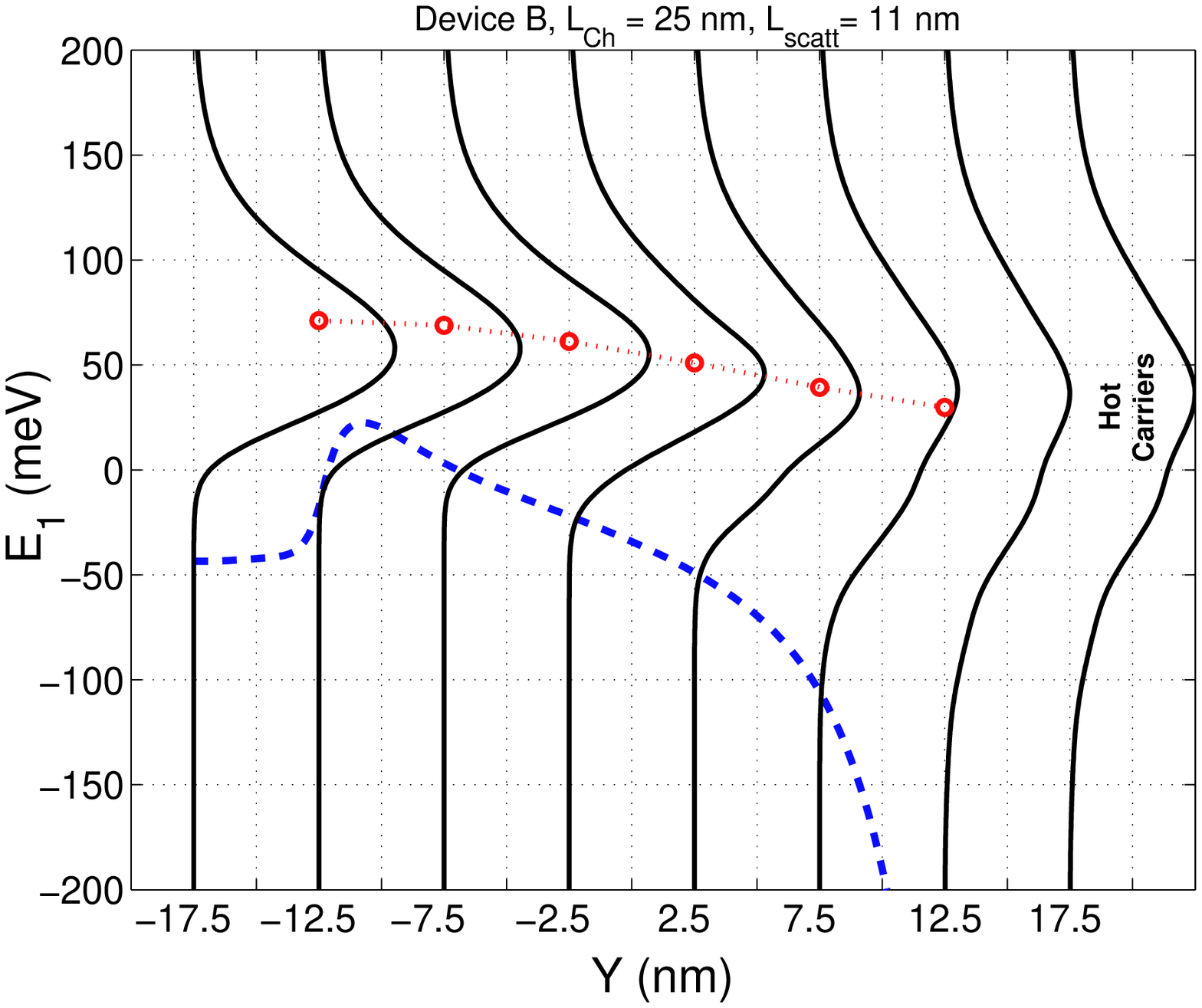}
     \epsfxsize=3.375in
\epsfbox{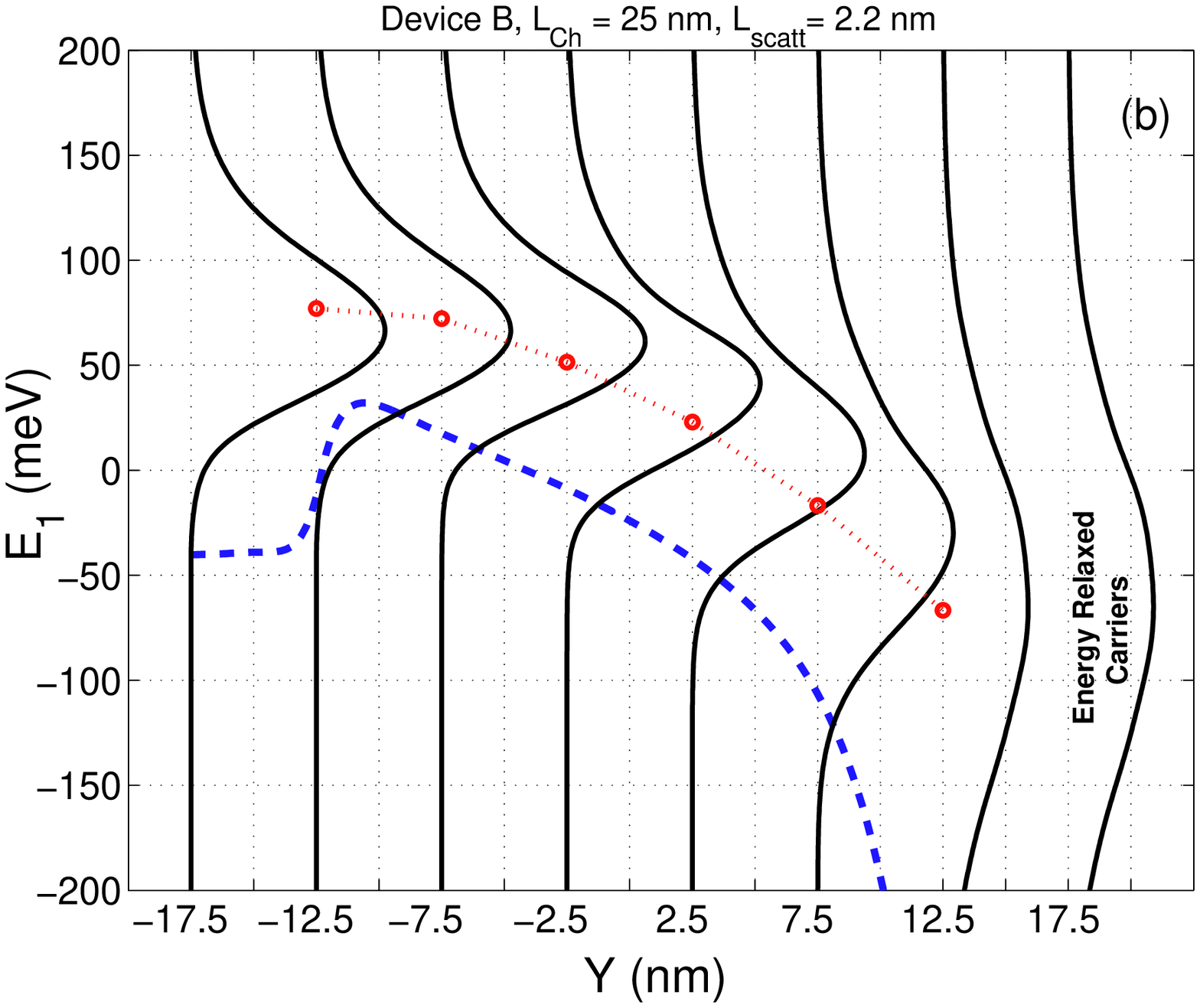}
  \end{center}
\vspace{2in}
\caption{\label{fig:paper_CURvsEvsY}}
\end{figure}

{\bf Fig. \ref{fig:paper_CURvsEvsY}:}
The solid lines represent $J(Y,E)$ for $Y$ equal to -17.5, -12.5, -7.5,
-2.5, 2.5, 7.5, 12.5 and 17.5 nm, from left to right respectively.
The dashed lines represent the first resonant level ($E_1$) along the
channel. The dotted lines represent the first moment of energy (mean)
with respect to the current distribution function $J(Y,E)$, which is
$\frac{\int dE E J(Y,E)}{\int dE J(Y,E)}$. (a) and (b) 
correspond to $L_{scatt} = 11$ and $2.2$ nm respectively in device B.
Scattering is included every where in the channel. (a) and (b) 
correspond to the $Y_{R-Scatt}=12.5$ nm data points of the dashed and
solid lines of Fig. \ref{fig:IdvsYRscatt25}. Scattering is
included everywhere in the channel but not in the extension regions. 
  
\vspace{0.5in}

\begin{figure}[htbp]
\begin{center}
     \leavevmode
     \epsfxsize=5.375in
\epsfbox{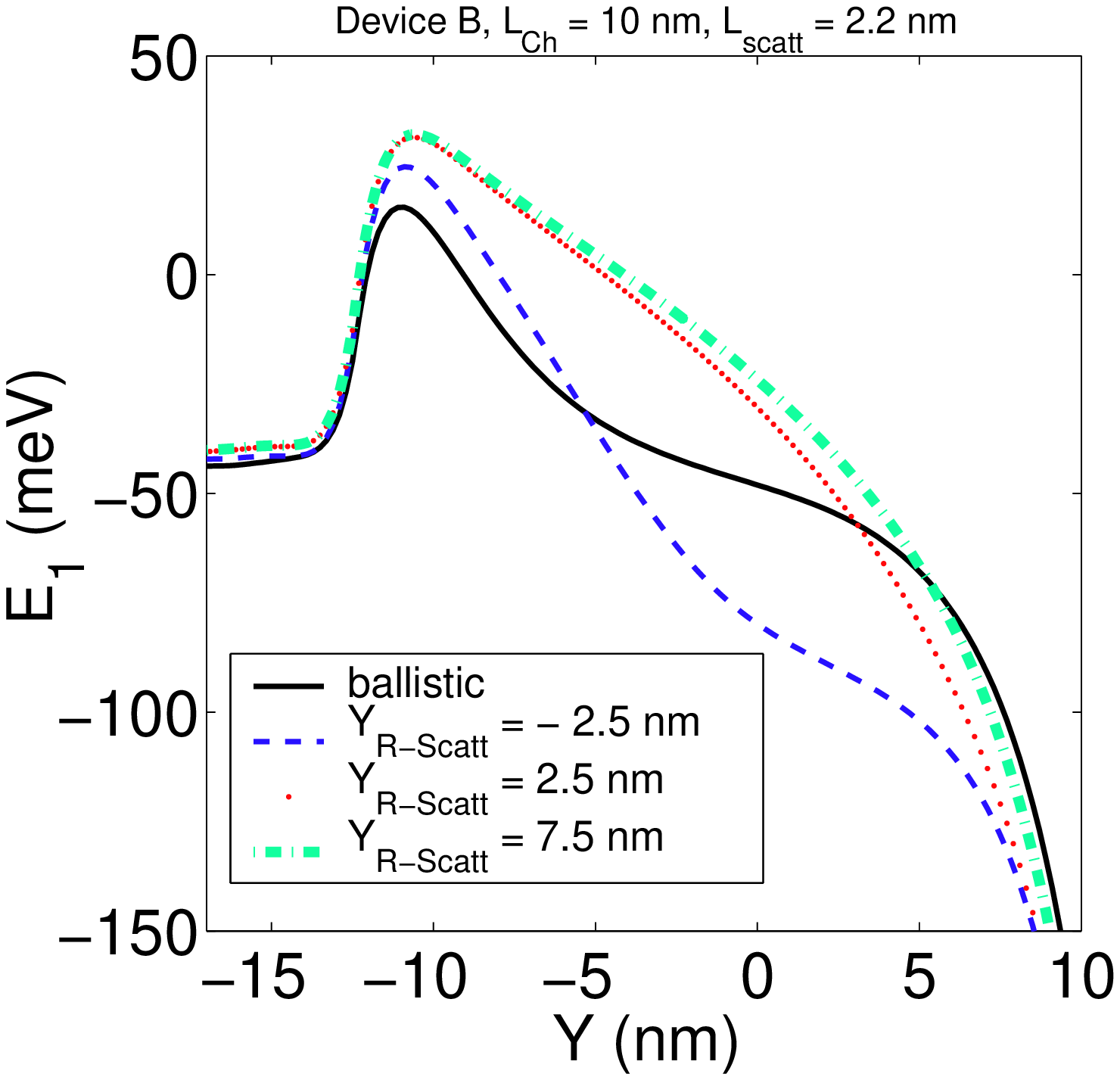}
  \end{center}
\vspace{2in}
\caption{ \label{fig:paper_POTvsY}}
\end{figure}

{\bf Fig. \ref{fig:paper_POTvsY}:}
Electrostatic potential versus $Y$ for device B. Scattering from -12.5
nm to 2.5 nm causes a large change in the source injection barrier 
($E_b$). Scattering to the right of 2.5nm causes a much smaller change
in $E_b$. In the absence of scattering, the potential profile in the 
channel tends to flatten. The potential drop (or $E_1$) along the 
channel is more ohmic / linear in the presence of scattering.

\vspace{0.5in}

\begin{figure}[htbp]
\begin{center}
     \leavevmode
     \epsfxsize=5.375in
\epsfbox{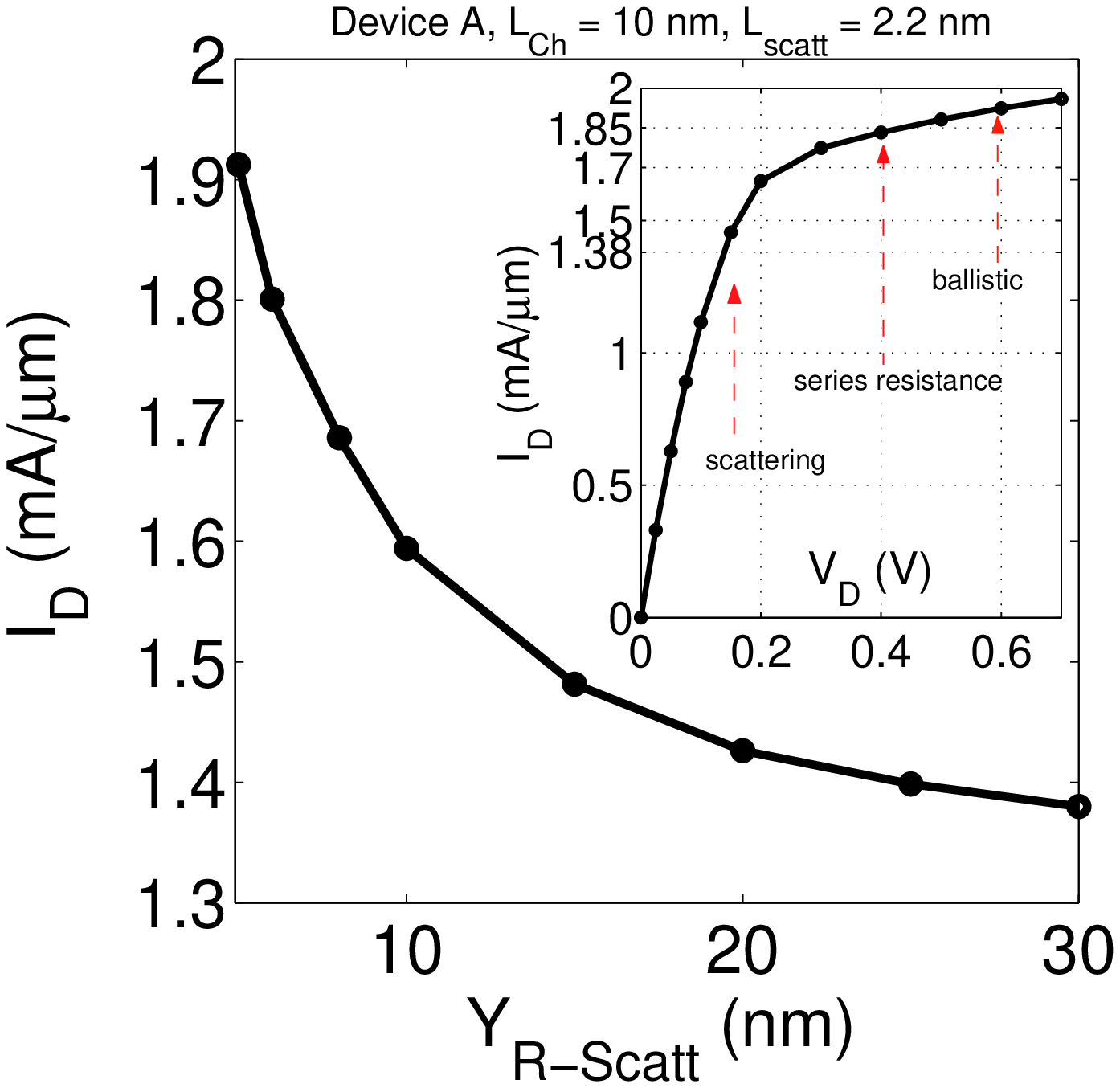}
  \end{center}
\vspace{2in}
\caption{ \label{fig:paper-series}}
\end{figure}

{\bf Fig. \ref{fig:paper-series}:}
$I_D$ versus $Y_{R-Scatt}$ for device A with scattering present 
only in the drain extension region from 5 nm to 30 nm. The large
reduction in drain current is due to scattering of hot carriers from
the drain extension region back in to the channel. The physics of this
effect is completely different from 'classical series resistance' in
MOSFETs, which is a much smaller effect. The results obtained from the
'series resistance' and 'scattering calculations' (this paper) are 
indicated by the arrows.
The electron-phonon scattering time is five times larger than in Fig.
\ref{fig:IdvsYRscatt10}. Inset: Drain current versus drain voltage
in the ballistic limit, showing the drain current estimate from the
series resistance picture and from our calculation.
%$I_D$ versus $Y_{R-Scatt}$ for devices A and C, with scattering present
%only in the drain extension region from 5 nm to 30 nm. The large
%reduction in drain current is due to scattering of hot carriers from
%the drain extension region back in to the channel. The physics of this 
%effect is completely different from classical series resistance in
%MOSFETs, which is a much smaller effect (see text). 
%The electron-phonon scattering time is five times larger than in Fig.
%\ref{fig:IdvsYRscatt10}. Inset: Drain current versus drain voltage
%in the ballistic limit for devices A and C. Device C has a much smaller
%DIBL than Device A. 

\pagebreak

\begin{table}[ht]
  \begin{center}
    \leavevmode
     \epsfxsize=5.375in
    \epsfbox{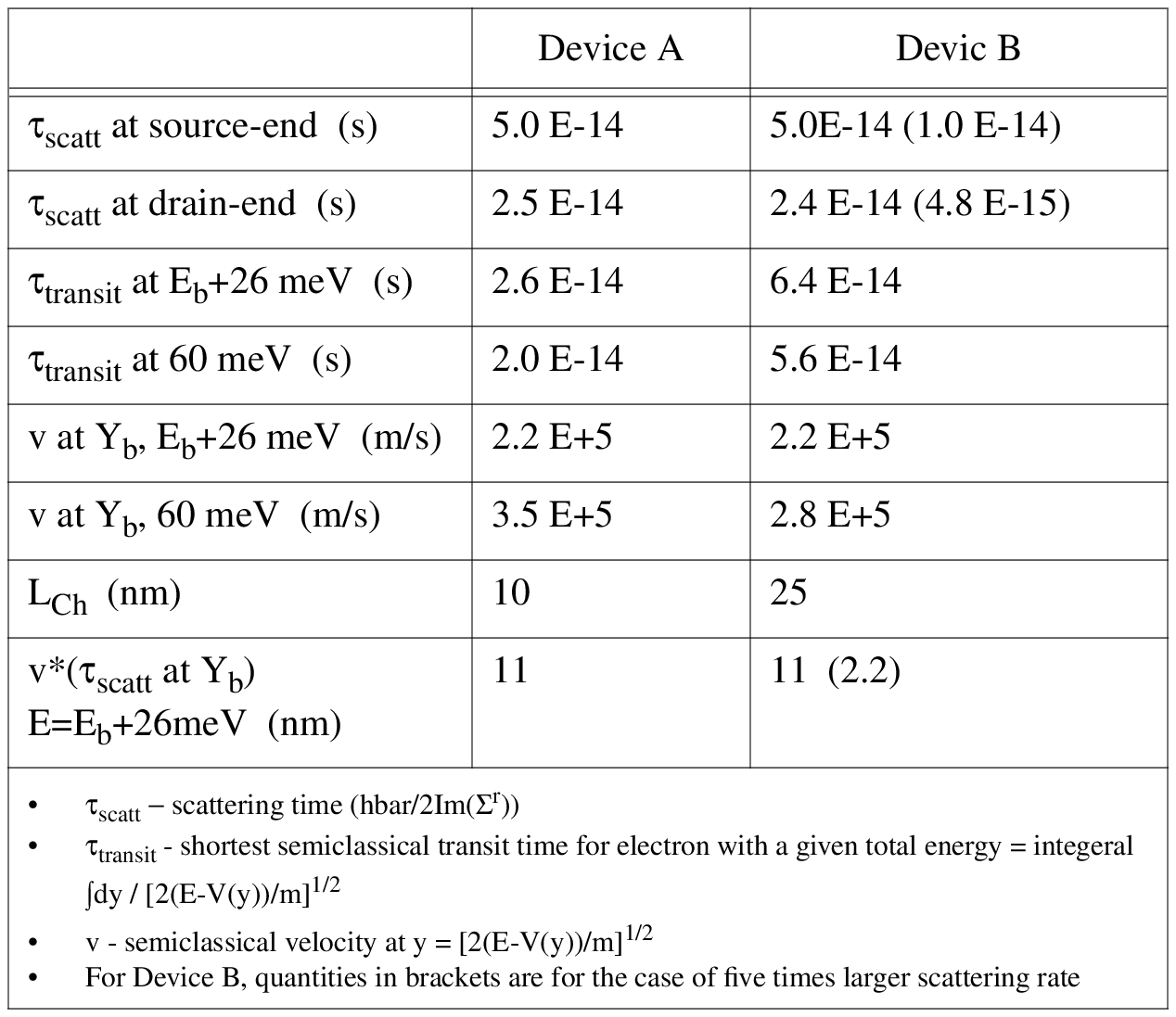}
  \end{center}
\caption{\label{table:time} Estimates of scattering time, transit time,
velocity and scattering length. }
\end{table}

\pagebreak

\pagebreak

\pagebreak

\pagebreak

\pagebreak

\pagebreak

%\pagebreak

\end{document}